\documentclass[10pt,journal,compsoc]{IEEEtran}

\ifCLASSOPTIONcompsoc
  \usepackage[nocompress]{cite}
\else
  \usepackage{cite}
\fi

\usepackage[ruled]{algorithm2e} 

\SetAlFnt{\small}
\SetAlCapFnt{\small}
\SetAlCapNameFnt{\small}
\SetAlCapHSkip{0pt}

\usepackage{amsmath}
\usepackage{algorithmic}
\usepackage{url}
\usepackage{hyperref}
\usepackage{graphicx}
\usepackage{amsfonts}
 \usepackage[table,xcdraw]{xcolor}
\usepackage{multirow}
\usepackage{microtype}

\begin{document}
%
\title{\huge 
Mapping Surfaces with Earcut}

%
%
%
%

\author{Marco~Livesu,~CNR IMATI}

\markboth{Journal of \LaTeX\ Class Files,~Vol.~14, No.~8, August~2015}%
{Shell \MakeLowercase{\textit{et al.}}: Bare Demo of IEEEtran.cls for Computer Society Journals}

\IEEEtitleabstractindextext{%
\begin{abstract}
Mapping a shape to some parametric domain is a fundamental tool in graphics and scientific computing. In practice, a map between two shapes is commonly represented by two meshes with same connectivity and different embedding. The standard approach is to input a meshing of one of the two domains plus a function that projects its boundary to the other domain, and then solve for the position of the interior vertices. 
Inspired by basic principles in mesh generation, in this paper we present the reader a novel point of view on mesh parameterization:
we consider connectivity as an additional unknown, and assume that our inputs are just two boundaries that enclose the domains we want to connect. We compute the map by simultaneously growing the same mesh inside both shapes.
This change in perspective allows us to recast the parameterization problem as a mesh generation problem, granting access to a wide set of mature tools that are typically not used in this setting.
Our practical outcome is a provably robust yet trivial to implement algorithm that maps any planar shape with simple topology to a homotopic domain that is weakly visible from an inner convex kernel. Furthermore, we speculate on a possible extension of the proposed ideas to volumetric maps, listing the major challenges that arise. Differently from prior methods -- for which we show that a volumetric extension is not possible -- our analysis leaves us reasoneable hopes that the robust generation of volumetric maps via compatible mesh generation could be obtained in the future.

\end{abstract}

}

\maketitle
\IEEEdisplaynontitleabstractindextext
\IEEEpeerreviewmaketitle


\newcommand{\cino} [1]{{\color{magenta}	Cino: #1}}
\newcommand{\com} [1]{{}} 
\newcommand{\edit} [1]{{\color{black}			#1}} 

\newcommand{\M}{M} 
\renewcommand{\l}{\ell} 
\renewcommand{\P}{P} 
\newcommand{\Q}{Q} 
\newcommand{\genus}[1]{{ g( #1 ) }}

\section{Introduction}
\label{sec:intro}


A one-to-one map between two topological spaces $A$ and $B$ is a function $f : A \leftrightarrow B$ that connects points in both domains. 
When it comes to actual coding, the realization of this mathematical idea is typically implemented using simplicial meshes to represent topological spaces.
Specifically, given two simplicial complexes $M_A,M_B$ that \emph{discretize} $A$ and $B$, the piece-wise linear map $f_M$ connecting them is implicitly defined by their shared connectivity.
If both meshes do not 
contain degenerate elements, boundaries do not self intersect and all triangles have coherent orientation, there exists a bijection $f_M : M_A \leftrightarrow M_B$ \cite{lipman2014bijective}. In fact, any point $p \in M_A$ is identified by a mesh element and a unique set of barycentric coordinates that locate $p$ inside it. Exploiting the shared connectivity, the image $\tilde{p} = f_M(p)$ can be located in $M_B$ by considering the same mesh element and the same barycentric coordinates. Switching $M_A$ with $M_B$ allows to navigate the map in the opposite direction.

\begin{figure}[t]
	\centering
	\includegraphics[width=\columnwidth]{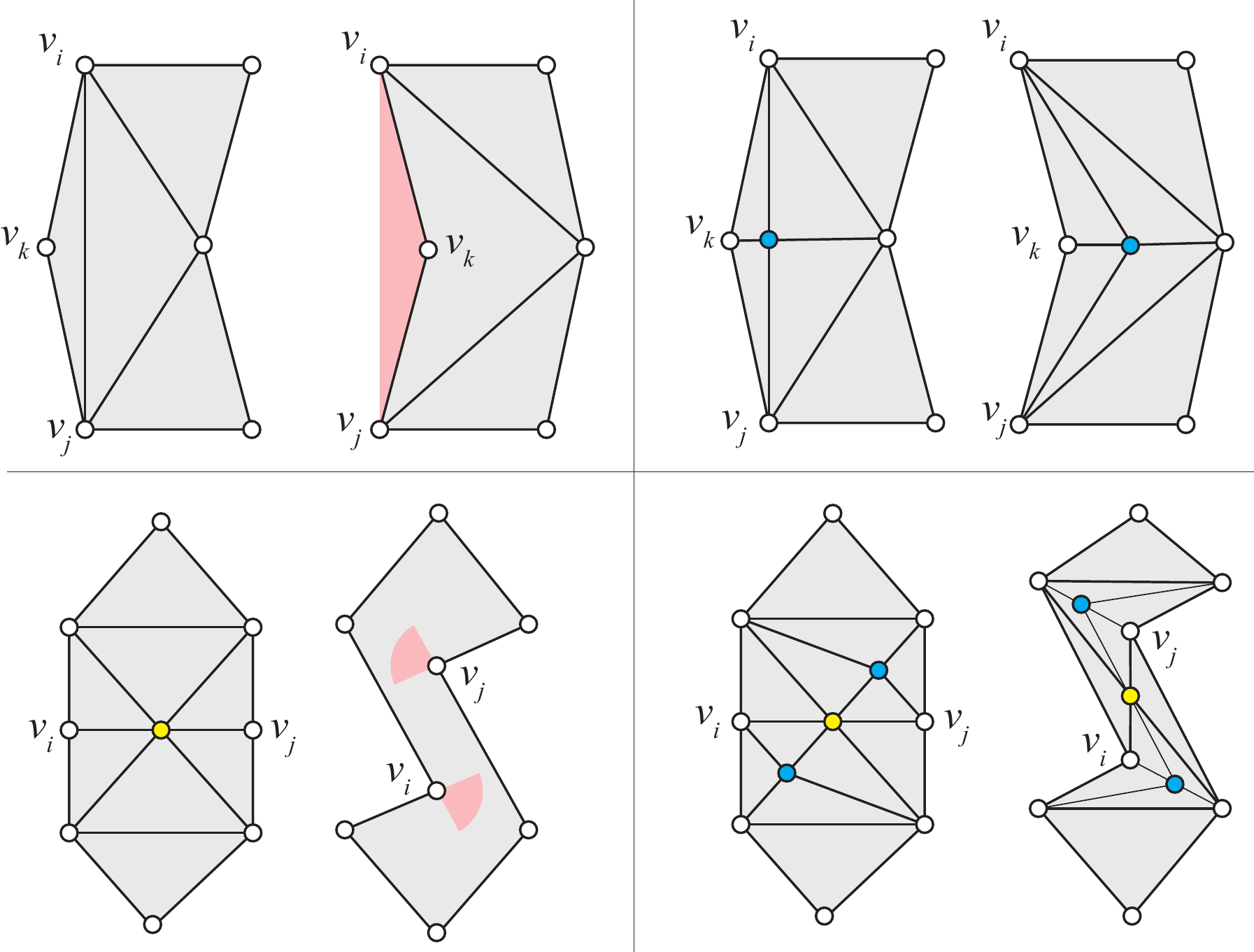}
	\caption{Top left: mapping the left polygon into the right one inevitably flips triangle $ijk$. Bottom left: similar deadlock configurations may also occur for interior vertices (yellow), because the intersection of the feasible cones of its neighbors is empty. Right column: inserting additional vertices in the domain (blue) allows to correctly complete both maps. Topological issues are hard to find with a static analysis of the input mesh, and may cause undefined behavior in the solver.}
	\label{fig:no_solution}
\end{figure}

Algorithms for the parameterization of a given mesh $M_A$ to some target domain $B$ typically input $M_A$ and the value of $f_M$ for each boundary vertex $p \in \partial M_A$, such that they jointly interpolate the boundary of $B$. Then, these methods internally compute $f_M$ for the interior vertices of the mesh, completing the embedding of $M_B$. The most widely used strategy to accomplish this task amounts to define some energy that encodes the distortion of the map, and then minimize it with numerical optimization. To this end, methods mainly differ for the types of energies they use and the numerical schemes used to minimize them. Despite hugely popular, methods based on numerical optimization suffer from two major problems: (i) they do not allow to edit the input topology, and may operate in an empty feasible space 
(Figure~\ref{fig:no_solution}); (ii) they often require the minimization of highly non linear energies, hence the output results depend on the initialization of the solver and its ability to land on a good local minimum.

In this paper we offer the reader a novel view on the mapping problem, unpinning it from the classical formulation based on numerical optimization. Keeping in mind that the ultimate goal is to produce two meshes that discretize the two domains and also share the same connectivity, we observe that an alternative way to formulate this problem consists in assuming as input two boundary representations of the spaces we want to connect, and to generate the wanted map by \emph{simultaneously growing the same mesh} inside both domains. Differently from prior methods, in this case the amount of interior vertices and the whole mesh connectivity are not fixed a priori, but rather become additional unknowns. This opens the space of solutions, allowing to design a mesh that is not tailored for one space only and then forcefully imposed in the other, but is rather a good compromise for both spaces (Figure~\ref{fig:no_solution}, right). 

We demonstrate our ideas in the context of planar maps, 
showing how the earcut triangulation algorithm can be used to initialize a provably bijective map between two polygons with simple topology. An earlier version of this paper was published at the international conference Smart Tools and Applications for Graphics, obtaining a honorable mention~\cite{Liv20b}. The proposed algorithm was limited to maps to strictly convex polygons. In this short paper we improve the original method, extending the class of admissible domains to any simple polygon that is weakly visible from a convex inner kernel~\cite{avis1981optimal,valentine1953minimal}. This class is much bigger than the class of convex polygons, and is also bigger than the class of star shaped polygons, which is fully included in it. 
We also conducted an extensive validation campaign, testing the new algorithm against more than two thousand shapes with varying complexity.


A major motivation for this research
stems form the fact that prior methods for robust 2D mapping do not extend to 3D. For the famous Tutte mapping~\cite{tutte1963draw} 
this was shown multiple times via counter examples~\cite{chilakamarri1995three,de2003tutte}. But even modern approaches such as Progressive Embeddings~\cite{shen2019progressive} raise major theoretical challenges going one dimension up (Section~\ref{sec:related}). 
In Section~\ref{sec:conclusions} we speculate on a possible extension of our findings to volume meshes, providing a list of (revised) computational geometry questions that form the theoretical foundation for tetrahedral meshing algorithms, but are still unanswered for the case of simultaneous meshing of two shapes at once. We believe that answering these questions will provide powerful tools to attack the volumetric mapping problem via mesh generation.

\section{Background}
\label{sec:related}

In this section we discuss methods for robust surface mappings that are closest to us, also motivating why they cannot be extended to volumes.

\begin{figure}
	\centering
	\includegraphics[width=.9\columnwidth]{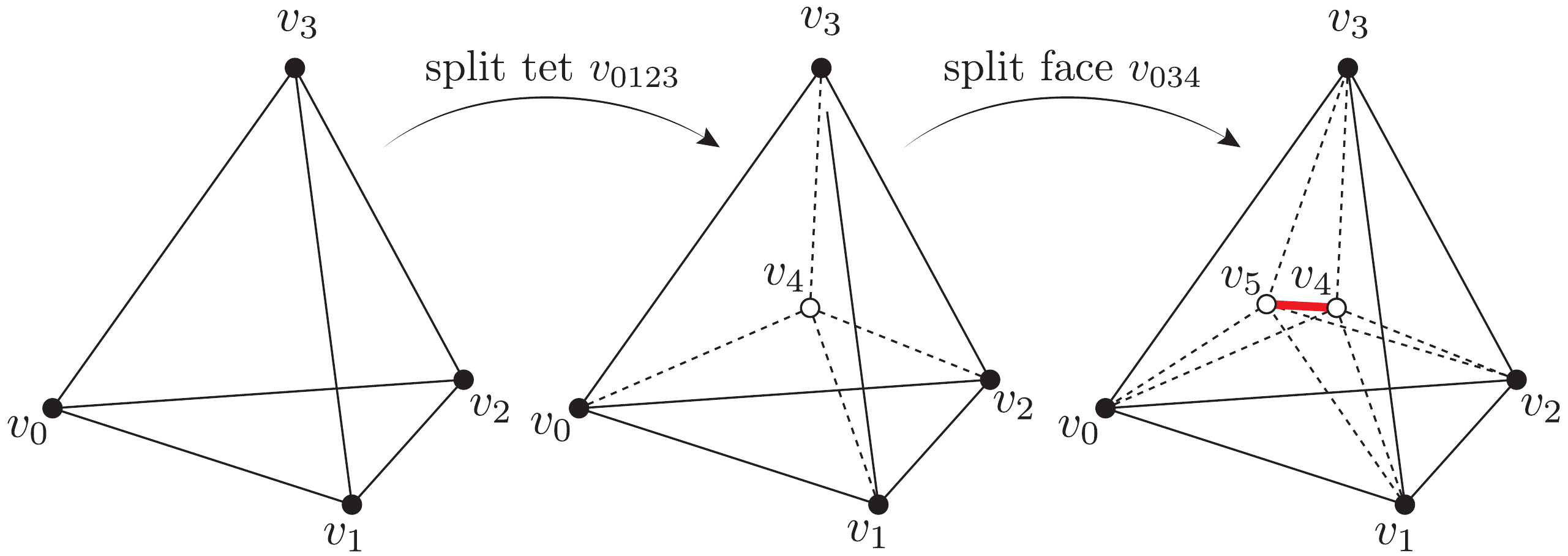}
	\caption{Splitting a tetrahedron into four sub tets, and then splitting any of the so generated interior faces yields a simplicial complex with four boundary vertices (black) and two inner vertices (white). Fixing the boundary and solving for the Tutte's barycentric mapping~\cite{tutte1963draw} produces a mesh where edge $v_4v_5$ (in red) is collapsed, thus breaking the map.}
	\label{fig:tutte3dfail}
\end{figure}

The Tutte embedding~\cite{tutte1963draw} was introduced in 1963 in the context of graph drawing, and was popularized in the graphics community by Floater in 1997, showing that any convex combination of neighbor vertices can be used to define the map~\cite{floater1997parametrization}. Ever since, a plethora of different methods have been proposed in the field, extending the original idea to topological tori and disk-like meshes with multiple boundaries~\cite{gortler2006discrete}, as well as porting it to other spaces, such as Euclidean~\cite{aigerman2015orbifold}, hyperbolic~\cite{aigerman2016hyperbolic} and spherical~\cite{aigerman2017spherical} orbifolds. Despite numerous attempts, it was shown multiple times that the barycentric mapping does not extend to 3d. This is shown with a concise counter example in Figure~\ref{fig:tutte3dfail}; other failure cases are reported in~\cite{chilakamarri1995three,de2003tutte}. It is thought that under some restricting assumptions a 3d extension could still be possible~\cite{chilakamarri1995three}, but we are not aware of any success in this regard.

\begin{figure}
	\centering
	\includegraphics[width=.9\columnwidth]{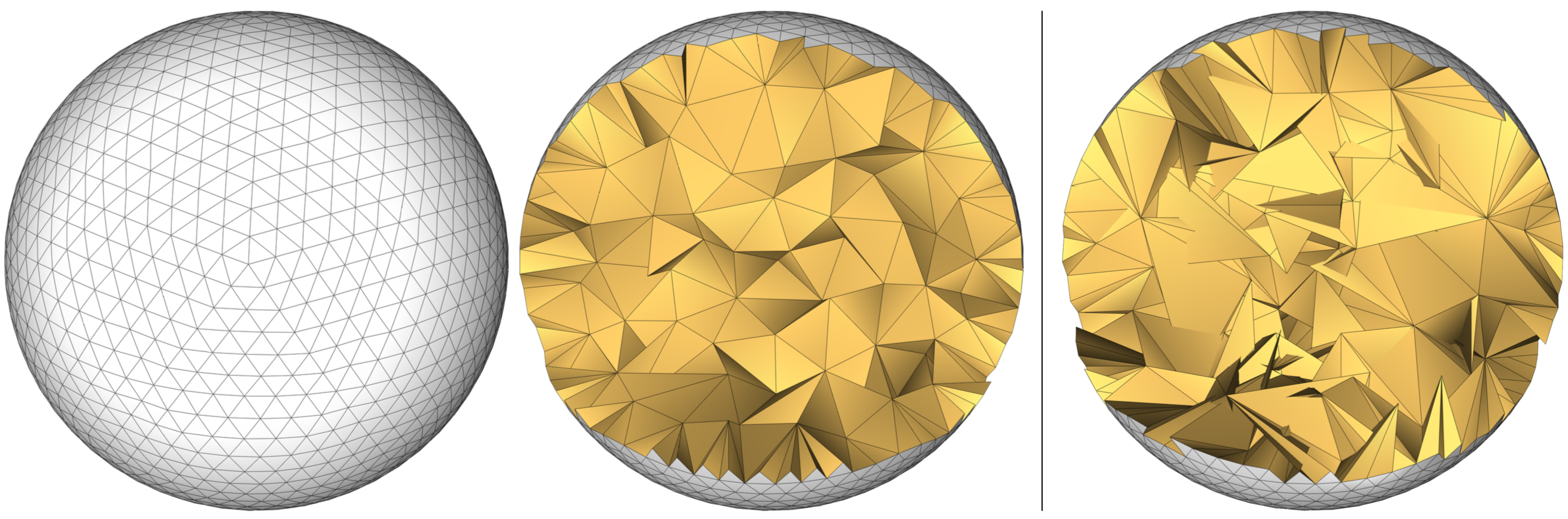}	
	\caption{Attempting to collapse the internal edges of a simple tetrahedralized sphere (left) along a random sequence always yields an incollapsible mesh where each edge violates the link condition~\cite{dey1998topology} (right). Applying barycentric subdivision allows to collapse new edges, but after each subdivision step the size of the incollapsible mesh we obtained was higher than the one at the iteration before.}
	\label{fig:incollapsiblesphere}
\end{figure}

The Tutte embedding offers theoretical guarantees but -- if pushed to the extreme -- concrete implementations may fail to produce a valid mapping because of the limited precision of floating point systems. Shen and colleagues propose an alternative method, called Progressive Embeddings~\cite{shen2019progressive}, which offers similar theoretical guarantees, but is less sensitive to floating point implementations. The algorithm is inspired by the progressive meshes concept~\cite{hoppe1996progressive}, and is based on the ability to deconstruct the topology of a triangle mesh by an ordered sequence of edge collapses, reconstructing the same mesh in another embedding with a sequence of vertex splits in the opposite order. Also this approach does not extend to 3d, the reason being twofold: (i) simplicial complexes in dimensions $d\geq 3$ may not be fully collapsible with a sequence of edge collapses, and even deciding whether a tetrahedral mesh is collapsible is NP Complete~\cite{tancer2016recognition,malgouyres2008determining,attali2014recognizing}. Theory says that after a finite set of barycentric subdivisions any simplicial complex becomes collapsible~\cite{adiprasito2019barycentric}, but still one should navigate the exponential space of all possible collapsing sequences to find a valid solution. Attempting to deconstruct a tetmesh along a heuristically computed collapsing sequence does not seem a good strategy either~\cite{lofano2019worst}. We personally tried many combinations of subdivisions and collapses, but always got stuck at some incollapsible configuration even on simple meshes (Figure~\ref{fig:incollapsiblesphere}); (ii) one may try to transform the input mesh into a mesh with different connectivity and known collapsing sequence via flip operators, but again this does not work because the graph of all possible triangulations of a given point set is connected for $d=2$~\cite{lawson1972transforming,osherovich2008all}, but there exist counter examples for the 3d case that show that it is disconnected for tetrahedralizations~\cite{dougherty2004unflippable}.
All in all, these issues basically kill any hope that similar ideas could be extended to tetrahedral meshes in a robust yet computationally feasible manner. 

\begin{figure*}[h]
	\centering	
	\includegraphics[width=\linewidth]{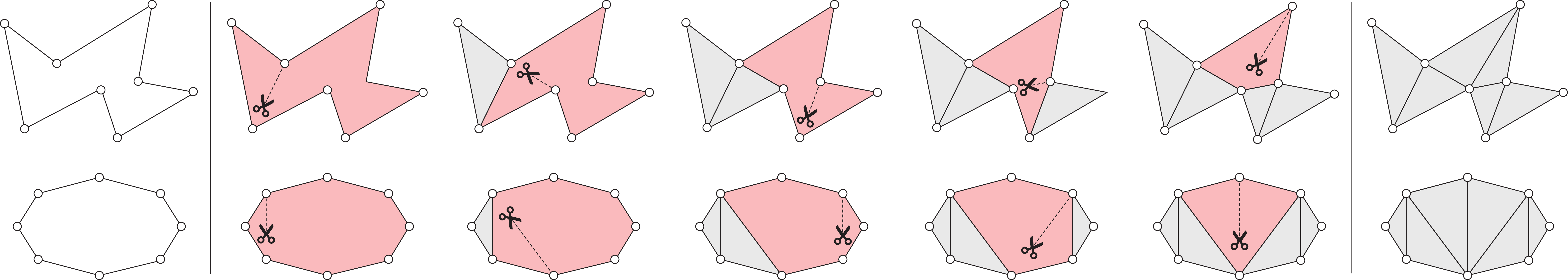}
	\caption{Our pipeline starts with two simple planar polygons (left). We the initialize two fronts (boundaries of the red shaded areas), which we progressively tessellate by adding triangles centered at front vertices that form valid ears (white circles) in both domains. Upon convergence, we obtain two meshes with same connectivity and different embedding (right).}
	\label{fig:example}
\end{figure*}

A variety of methods offer the ability to perform a cross parameterization between two surface meshes of same topology. They employ composition of maps to intermediate domains such as coarse base complexes~\cite{kraevoy2004cross,schreiner2004inter} or some polygonal schema~\cite{weber2014locally,Liv20,yang2020error}. These methods internally employ standard techniques (e.g. Tutte, or some derivation of it) to generate the underlying maps, therefore could potentially benefit from our contribution, and are orthogonal to it. 

Volume methods 
formulate the mapping as an optimization problem. As for the 2d case, topology is fixed in input, and differences arise in the energies and numerical schemes used. We count mainly two family of approaches: (i) methods that input an invalid map and project it into the feasible space by fixing inverted elements~\cite{aigerman2013injective,kovalsky2014controlling,su2019practical,du2020lifting}; (ii) methods that input a valid solution and iteratively reduce geometric distortion minimizing barrier energies that grow to infinity if an element becomes nearly degenerate or flips its orientation~\cite{rabinovich2017scalable}. The former do not provide any guarantee, and may fail to even enter the feasible space. The latter guarantee the generation of a valid map if correctly initialized. In 2d the initialization step can be computed with~\cite{tutte1963draw,shen2019progressive}. We are not aware of any 3d method that can provably generate a valid initial solution. 

A simplified version of the volume mapping problem was proposed in~\cite{campen2016bijective}. This method is based on foliations, and allows to map a genus zero simplicial mesh to a cube or a sphere. Users can select the type of foliation (e.g. radial for the mapping to a sphere) but cannot prescribe point to point correspondences for surface vertices. Moreover, generating a valid piece-wise linear map requires to perform aggressive mesh refinement, greatly increasing output mesh size even for simple objects.


\section{The Earcut Mapping Algorithm}
\label{sec:method}

In this section we introduce our novel surface mapping algorithm. We input two boundary representations of the domains $A$ and $B$ we want to connect in the form of two closed chains of vertices.
Domain $A$ is a simply connected planar region, possibly containing concavities. Domain $B$ must be strictly convex.
If both chains do not contain vanishing edges there exists a one to one map $f_\partial : \partial A \leftrightarrow \partial B$. We extend such map from the boundary to the interior, providing in output a piece-wise linear function $f_M : M_A \leftrightarrow M_B$ in the form of two triangle meshes $M_A,M_B$ with same connectivity.


Our key ingredient is a reformulation of the mapping problem in terms of mesh generation. We take inspiration from advancing front meshing algorithms such as~\cite{lohner1996extensions,marcum1995unstructured}, which start from a boundary (or initial front) describing an empty region to be meshed, and obtain the output mesh by progressively attaching new elements to such front, until it completely vanishes. Our key observation is that if we consider as initial fronts two regions we want to connect, and we define a sequence of advancing moves that are compatible with both fronts, we will obtain the same meshing for the two domains, hence a mapping between them. Compared to classical approaches, our special setting imposes three important differences:
\begin{itemize}
\item[1)] we work simultaneously in two domains, therefore advancing moves must be valid in both fronts, and must be applied following the same order;
\item[2)] at any time during execution, there must be a one to one correspondence between fronts in both domains, which must therefore be homotopic and contain the same number of vertices and edges;
\item[3)] assuming the absence of degenerate or flipped elements, \emph{any} meshing is ok, regardless of the quality of its elements. This differentiates from classical mesh generation algorithms, which largely concern about per element quality
\end{itemize}

The first condition ensures that all mesh elements we introduce have their own linear map connecting their two copies in both domains. The second condition ensures that the algorithm does not get stuck by creating topological mismatches in the fronts, which would prevent the completion of a valid global map. 
The third condition is just optional, but certainly applies to our case: we are only interested in initializing a valid map, without caring about geometric distortion. Methods that wish to generate a low distortion map may also leverage techniques for high quality advancing front mesh generation, but this is outside of the scope of this paper.

The algorithm works as follows: we initialize two fronts $F_A,F_B$ with the two input chains of vertices $A,B$. Since each domain is a simply connected polygon, we can use trivial earcut~\cite{eberly2008triangulation} to advance the front. Specifically, we detect convex front vertices in $F_A$ (i.e. vertices having inner angle lower than $\pi$), and check whether the triangle they form with their left and right neighbors contains any other vertex of $F_A$. If this is not the case, it means that the triangle is a valid ear, which can be \emph{cut} (i.e. removed) from the front, yielding a simpler polygon with one vertex less. Whenever a valid ear is found, the corresponding vertex is removed from both $F_A$ and $F_B$.
Note that by our initial assumption $F_B$ is convex, hence any of its vertices forms a valid ear. Moreover, any ear cut from it will preserve its convexity, because removing a point from a convex polygon yields a simpler convex polygon. This greatly simplifies the meshing process, because it allows us to produce two meshes simultaneously by caring on the validity of each move only in one of the fronts ($F_A$). The iterative process continues as long as the size of the fronts is greater than 3. Once $\vert F_A \vert = \vert F_B \vert = 3$, we can complete the meshing by adding a triangle lid that vanishes both fronts. An algorithmic description of the method is given in Algorithm~\ref{alg}. Figure~\ref{fig:example} shows all the iterations for a simple example. 

\textbf{Convergence.} Decisions on which ear should be cut are always taken w.r.t. a single front ($F_A$). The problem is therefore equivalent to triangulating domain $A$, and convergence is guaranteed by the fact that any simple polygon has at least two valid ears~\cite{meisters1975polygons}, which is also the theoretical foundation for the earcut algorithm we employed. 

\textbf{Complexity.} The complexity of the algorithm fully depends on the triangulation method of choice. For simplicity and ease of reproduction we opted for earcut. 
Despite optimal for a certain family of shapes~\cite{livesu2020deterministic}, the asymptotic complexity of this method for general simple polygons is
 $\mathcal{O}(n^2)$~\cite{elgindy1993slicing,eberly2008triangulation}. Note that the mapping paradigm we propose is not strictly linked to earcut, which could be virtually substituted with any other triangulation algorithm, obtaining a different asymptotic complexity.


\begin{algorithm}[t]
	\textbf{Input:} two closed lists of vertices $V_A, V_B$, such that both chains form simple polygons, $V_B$ is convex, and $\vert V_A \vert = \vert V_B \vert$.\\
ng a bijection $f_M : M_A \leftrightarrow M_B$\\
	\hrulefill \\
	\SetAlgoLined
	\vspace{0.1em}
	$M_A = (V_A, \emptyset)$\;
	$M_B = (V_B, \emptyset)$\;
	initialize front $F_A$ with $V_A$\;
	initialize front $F_B$ with $V_B$\;
	\While{$\vert F_A \vert > 3$ }
	{
		find an index $i$ such that triangle $v_{i-1}, v_i, v_{i+1}$ is a valid ear in $F_A$\;
		insert triangle centered at $i$ in both $M_A$ and $M_B$\;
		$F_A = F_A \setminus i$\;
		$F_B = F_B \setminus i$\;
	}
	fill the triangular hole in $M_A$ with verts in $F_A$\;
	fill the triangular hole in $M_B$ with verts in $F_B$\;
	\textbf{return $M_A, M_B$}\;
	\caption{}
	\label{alg}
\end{algorithm}

\section{Extension to weakly visible polygons}
\label{sec:extension}
\begin{figure}
	\centering
	\includegraphics[width=.9\columnwidth]{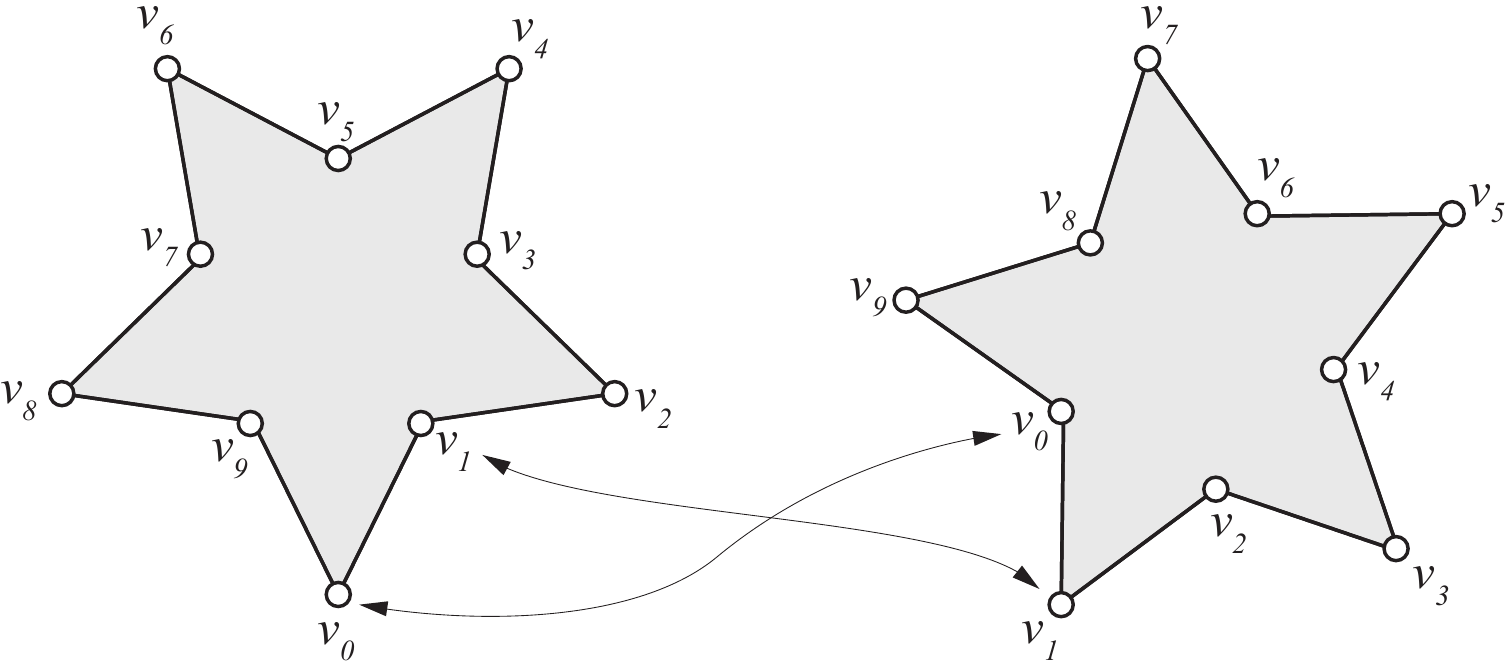}
	\caption{Failure case for Algorithm~\ref{alg}: the two polygons cannot be mapped to one another because convex vertices in the left star correspond to concave vertices in the right one, and vice versa. A mapping between the two stars is still possible if additional points are permitted, e.g. with the offsetting technique described in Section~\ref{sec:extension}.}
	\label{fig:deadlock2d}
\end{figure}

The algorithm presented in Section~\ref{sec:method} can only be used if one of the two domains is strictly convex. Mappings to non strictly convex polygons (e.g. a square) are not supported because if a triangle maps to three co-linear points it becomes degenerate. Similarly, concave polygons are not supported because they would require to validate each ear in both domains, possibly leading to deadlock configurations (Figure~\ref{fig:deadlock2d}).
In this section we discuss an extension of the mapping algorithm that permits mappings to any polygon that is weakly visible from an inner strictly convex domain~\cite{avis1981optimal,valentine1953minimal}. 

The basic idea is to split the mapping process in two parts. We first create an offsetting of the two domains, generating two sub polygons that live inside the input ones. One offset is generated with a topological approach, and can produce simple polygons of any shape. The other offset is computed with a geometric approach, and is guaranteed to produce a strictly convex polygon. Then, the mapping between offset polygons is generated with Algorithm~\ref{alg}, whereas the space in between offsets and the outer polygons is triangulated separately. In the remainder of this section we detail all the technical aspects of this modified pipeline.


\begin{figure*}
	\centering
	\includegraphics[width=\linewidth]{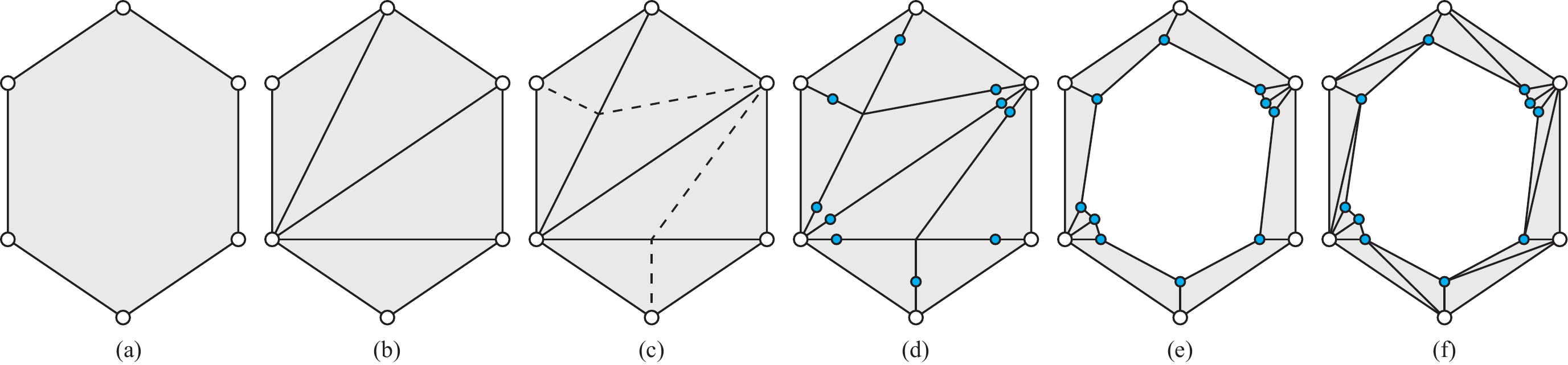}
	\caption{Pipeline for our robust topological offsetting: (a) input polygon; (b) a triangulation of it; (c) splitting edges to ensure that each vertex has at least one incident edge not on the boundary; (d) adding offset points by splitting inner edges incident to each input vertex; (e) connecting the new vertices to form the inner chain; (f) completing the triangulation by adding quad diagonals. Exploiting the underlying mesh we can guarantee that the resulting offset does not self intersect, and that all formed triangles have consistent orientation.}
	\label{fig:offsetting}
\end{figure*}

\begin{figure*}
	\centering
	\includegraphics[width=\linewidth]{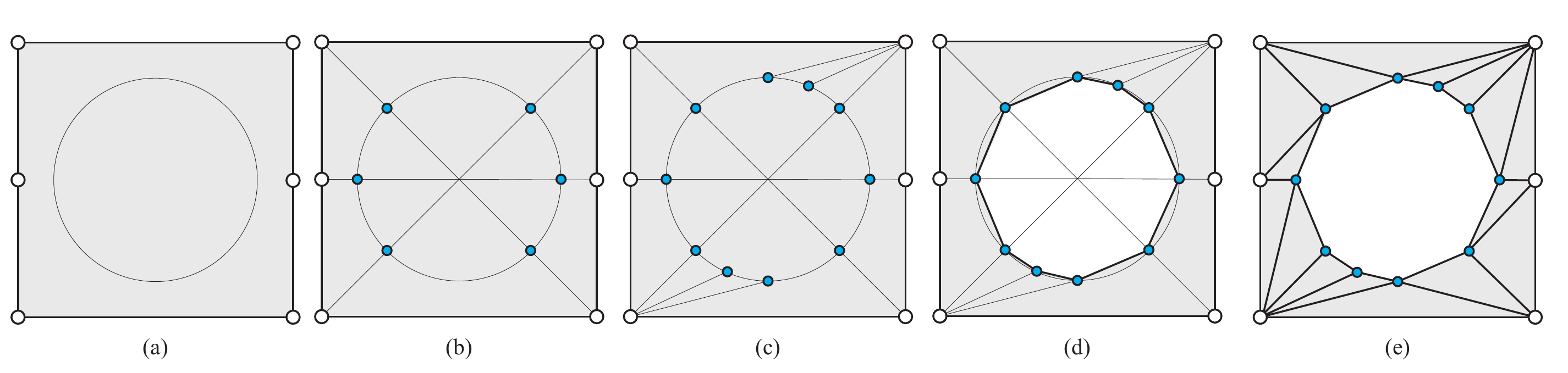}
	\caption{Mapping the offsetted polygon shown in Figure~\ref{fig:offsetting} to a square: (a) we initialize a circle inside the square; (b) each outer point is mapped to the circle through the line connecting it with the circle center; (c) outer points that had multiple inner points introduce new samples in the circle, thus preserving strict convexity; (d) inner points are connected to form the offset polygon; (e) the space in between outer and inner polygons is triangulated.}
	\label{fig:offset_mapping}
\end{figure*}

\textbf{Topological offsetting.} Given a simple polygon $P$ with $n>3$ vertices, the topological offsetting produces a simple polygon $P'$ with $m \geq n$ vertices that is strictly contained in it. The offsetting procedure is depicted in Figure~\ref{fig:offsetting}: we firstly triangulate $P$, and then refine the so generated mesh in order to make sure that for each point $p \in P$ there is at least one incident edge that is not on the boundary. Since any simple polygon contains at least two valid ears~\cite{meisters1975polygons}, there will be at least two vertices with only two incident edges that require mesh refinement. Specifically, it can be proved that $\vert P' \vert \geq \vert P \vert +2$ for any polygon with five or more vertices. For each boundary vertex $p \in P$ we introduce as many offset points as the number of incident inner edges. We eventually connect these points in a closed loop, generating the wanted offset polygon $P'$. Working on an underlying triangulation gives great robustness to this method, which provides three important guarantees: (i) offset points are always positioned strictly inside the input polygon, because they sample inner edges in the triangulation; (ii) the offset polygon is topologically simple and forms a closed chain of vertices. Consistent vertex ordering is guaranteed by the underlying mesh topology; (iii) the offset polygon does not self intersect, because its sides can all be formed by splitting edges in the underlying triangulation.

\begin{figure}
	\centering
	\includegraphics[width=.9\linewidth]{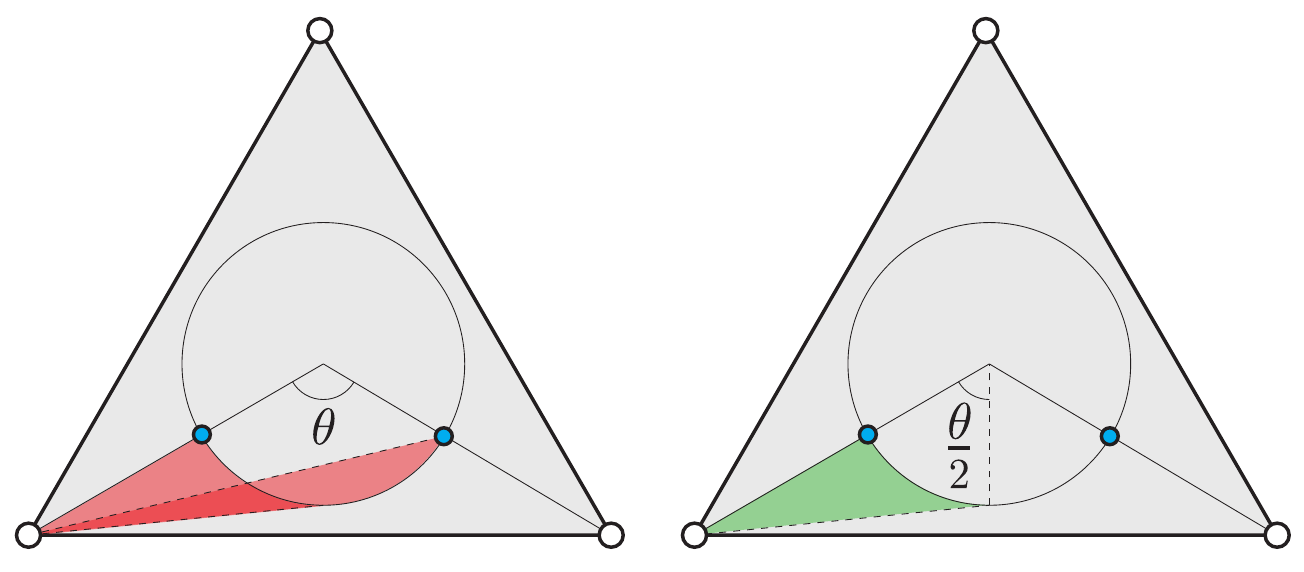}
	\caption{Inserting additional vertices in the inner circle interpolating the angle between adjacent points may produce fold overs if the whole angle extent is used (left). Considering only half of the available angle allows to easily and robustly insert all the necessary points (right).}
	\label{fig:offset_half_angle}
\end{figure}

\textbf{Geometric Offsetting.} The goal of this step is to reproduce the polygon obtained with topological offsetting in the second domain, ensuring that the result is a strictly convex polygon. 
The geometric offsetting algorithm inputs a polygon $P$ and a disk $D \subset P$ with radius $r$ and center $c$. We assume that any corner of $P$ sees a portion of $D$, including its center. In other words, $D$ is \emph{weakly visible} from $P$~\cite{avis1981optimal,valentine1953minimal}. Note that weak visibility is a less restricting assumption than star shapedeness, hence $D$ must no be strictly contained in the kernel of $P$, although our geometric construction requires that at least its center $c$ does.
The first step of the algorithm is to map each corner of $P$ onto $D$, computing the intersection between $D$ itself and the line connecting the corner with center $c$. These points already define a convex offset polygon with size $\vert P \vert$. However, topological offsetting generates polygons with bigger size, 
therefore we need to sample disk $D$ at some extra point to ensure that both offsets have the same number of vertices and same ordering. Let $p_i,p_{i+1} \in P$ be two adjacent points in the outer polygon, and $p_i',p_{i+1}' \in D$ the two offset points associated to them. We assign to point $p_i$ the circular arc delimited by $p_i',p_{i+1}'$, and use it to sample additional offset points that may be needed to match the topological offset obtained in the other domain. Additional points are generated so as to uniformly sample the first half of the arc. We leave the second half of the arc empty in order to avoid flips in the triangulation (Figure~\ref{fig:offset_half_angle}). Once all necessary points have been added, the geometric offset will contain the wanted number of vertices and be strictly convex, because all its points have been sampled on the input disk $D$.

\textbf{Triangulation.} We compute a compatible triangulation of the space in between the outer polygon and its inner offset as follows.
Let us focus on polygon $A$ and its offset $A'$; the same triangulation scheme applies to $B,B'$ as well. Let $p_i \in A$ a vertex in the outer polygon, and $p'_{i,1},\dots,p'_{i,n} \in A'$ the offset points that are associated to it. We first generate $n-1$ triangles connecting adjacent offset vertices with $p_i$, forming triangles $p_i,p'_{i,j},p'_{i,j+1}$ for any $j = 1,\dots,n-1$. Then, we triangulate the remaining quads formed by vertices $p_i,p_{i+1},p'_{i,n},p'_{i+1,1}$ by tracing the diagonals $p_{i+1},p'_{i,n}$. The result of this triangulation can bee seen at the rightmost column of Figures~\ref{fig:offsetting} and~\ref{fig:offset_mapping}. We eventually complete the map by exploiting the fact that offset $B'$ is a strictly convex polygon, hence we run Algorithm~\ref{alg} using offsets $A'$ and $B'$ as input. 

\begin{figure}
\centering
\includegraphics[width=.9\columnwidth]{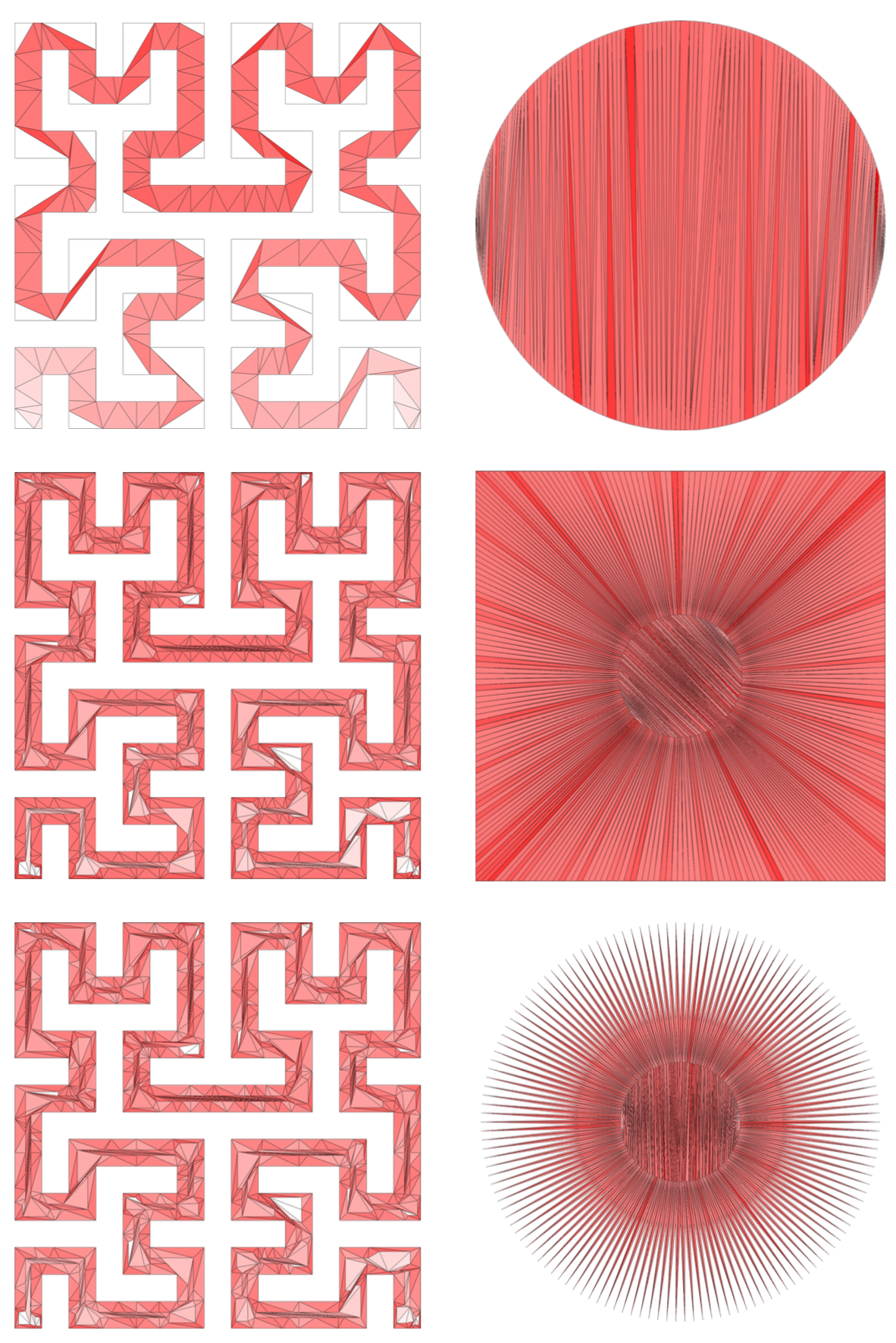}
\caption{Mapping of a space filling curve to a circular (top), squared (middle) and star-shaped (bottom) domain. Triangles are colored from white to red according to the logarithm of their $\ell_2$ stretch~\cite{sander2001texture}.}
\label{fig:space_filling}
\end{figure}

\begin{figure}
	\centering
	\includegraphics[width=.19\columnwidth]{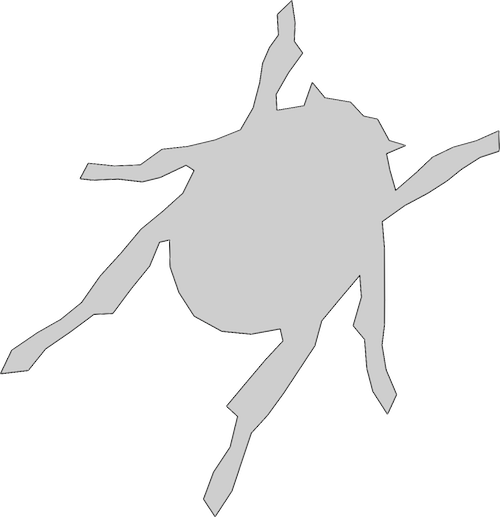}
	\includegraphics[width=.19\columnwidth]{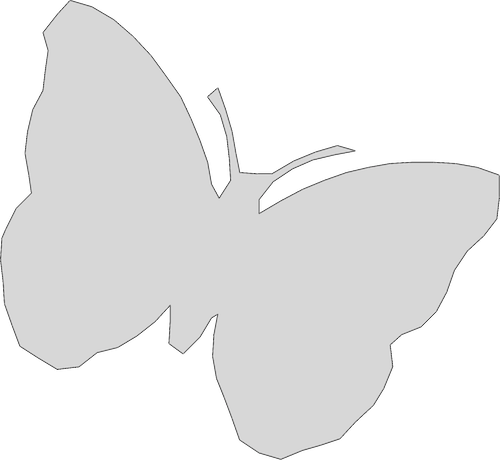}
	\includegraphics[width=.19\columnwidth]{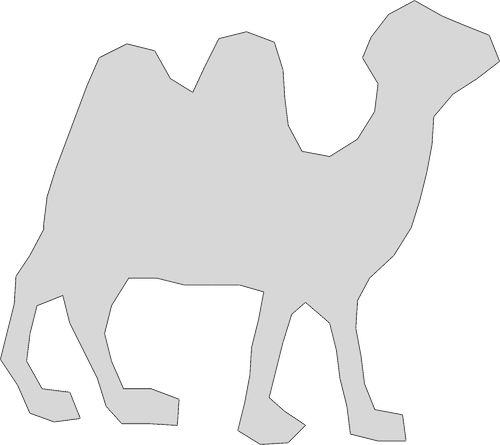}
	\includegraphics[width=.19\columnwidth]{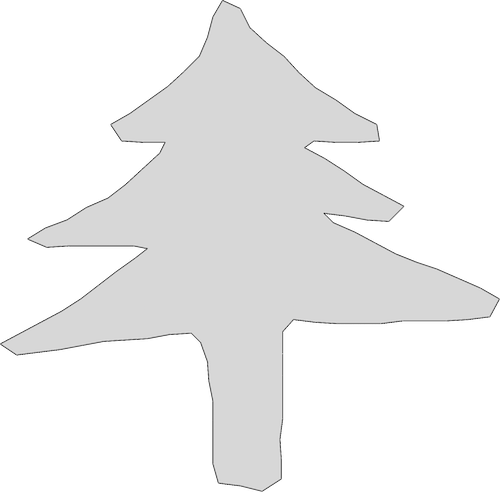}
	\includegraphics[width=.19\columnwidth]{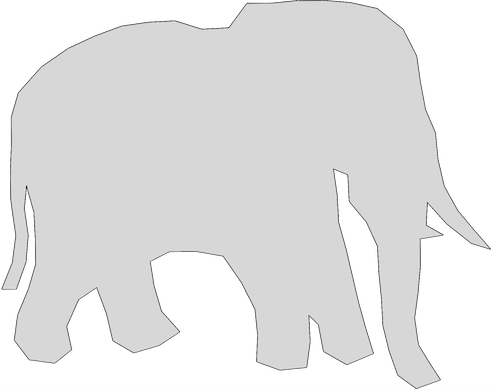}\\
	\includegraphics[width=.19\columnwidth]{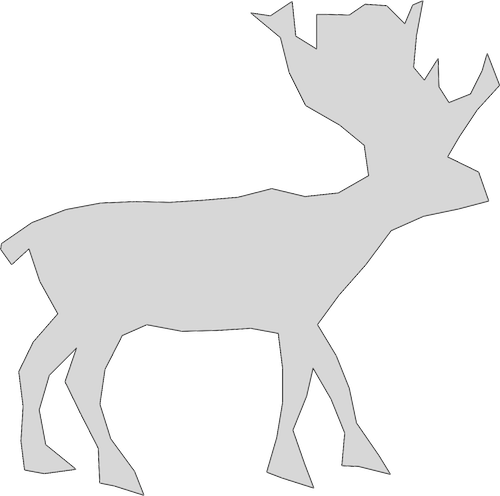}
	\includegraphics[width=.19\columnwidth]{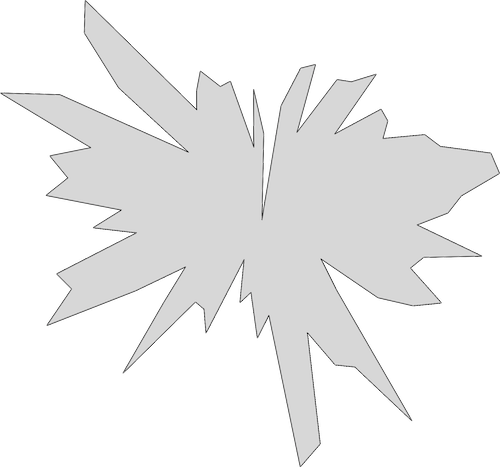}
	\includegraphics[width=.19\columnwidth]{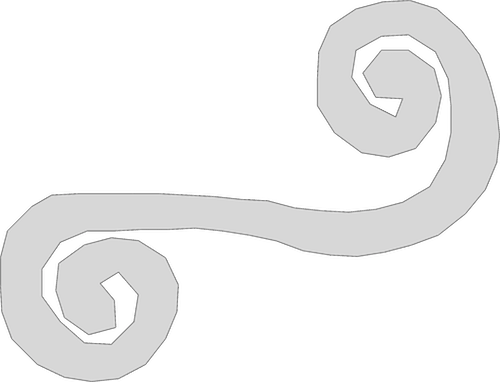}
	\includegraphics[width=.19\columnwidth]{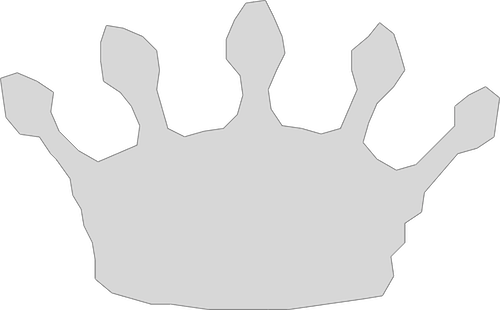}
	\includegraphics[width=.19\columnwidth]{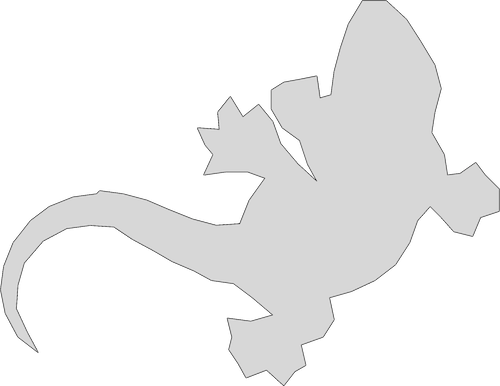}
	\caption{A sampling of the 2K shapes used in our experiments. For each model we produced three mappings: to a circle, a square and a star.}
	\label{fig:dataset}
\end{figure}

\section{Validation}
We implemented the earcut mapping and the offset algorithms in C++, and publicly released them as a module of CinoLib~\cite{cinolib}. We tested our algorithms on a MacBook Pro with an i5 2.9GHz processor and 16GB of RAM. For our tests, we collected a large dataset of shapes which comprises 1039 contours from the 2D shape structure dataset~\cite{carlier20162d}, and 1000 splat-like random polygons procedurally generated using Blender~\cite{blender} (Figure~\ref{fig:dataset}). For each shape in our benchmark we produced three alternative mappings: to a circle, a square, and a star-shaped domain (Figure~\ref{fig:space_filling}). Overall, we counted 2039 different shapes, and 6117 mappings. In all cases, our algorithms produced a valid (i.e. bijective) map, which we verified using exact orientation predicates~\cite{shewchuk1997adaptive}. Maps to circles involve only the earcut mapping Algorithm~\ref{alg}, and took 0.4 seconds overall. Maps to squares and stars also require topological and geometric offsetting, and took 3.77 and 3.68 seconds, respectively.

\begin{figure}
\centering
\includegraphics[width=\columnwidth]{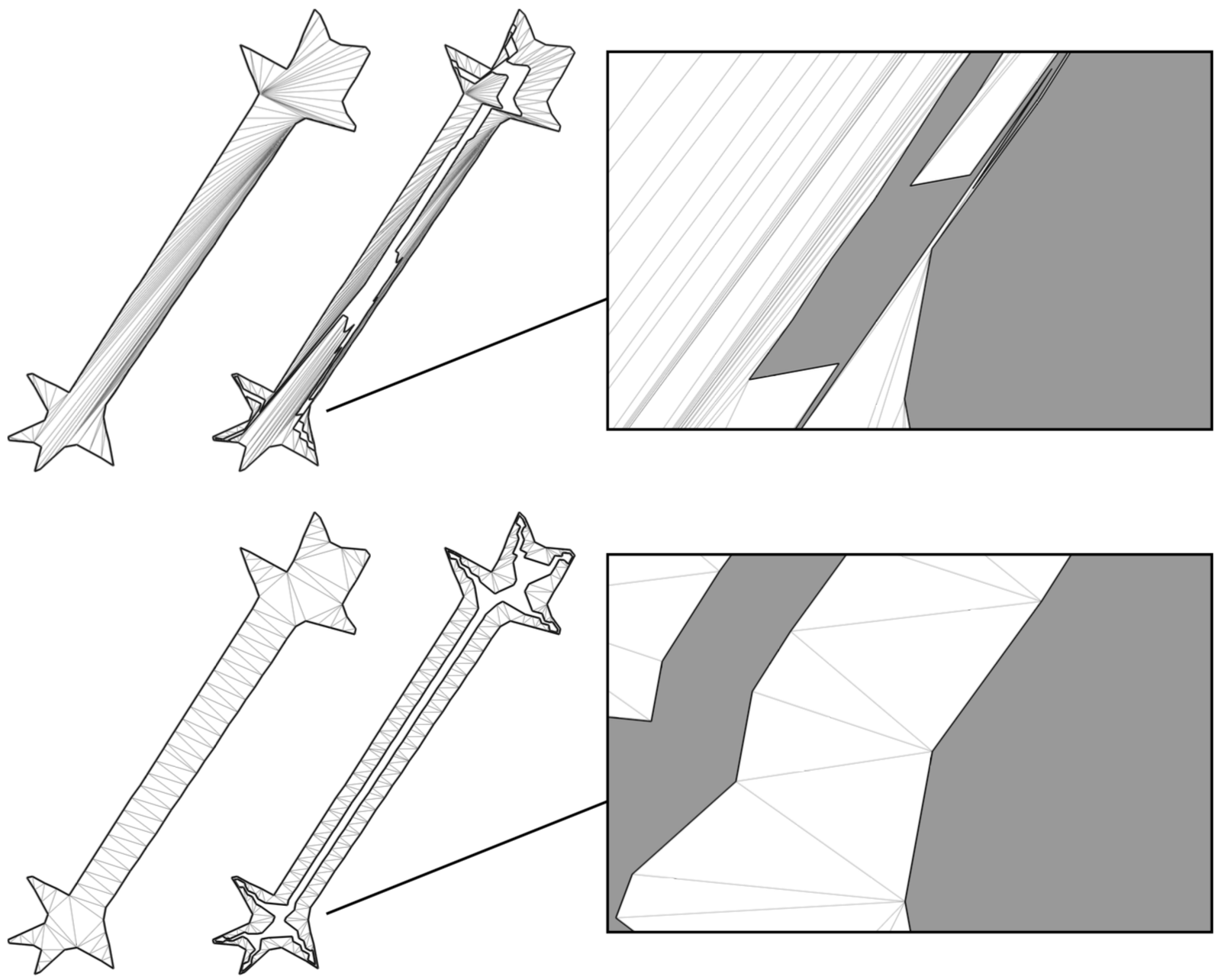}
\caption{Top: topological offsetting is sensitive to extremely low quality guiding meshes, because offset points sample the edges of nearly degenerate triangles, possibly introducing degenerate or flipped elements due to numerical roundoff errors. Bottom: using a higher quality triangulation makes the algorithm numerically stable. In practice, we substituted classical earcut, which processes ears in the order they are found (top), with a prioritized earcut, which orders ears according to their angle (bottom).}
\label{fig:earcut}
\end{figure}

These experiments were particularly useful to give us insight on the  practical robustness of a floating point implementation. We soon observed that if the triangulation used to guide the topological offsetting contained elongated -- nearly degenerate -- triangles, then issues related with numerical precision could occasionally introduce flipped elements (Figure~\ref{fig:earcut}). This happened on approximately 100 mappings out of 6117.
From a mesh generation perspective this is not surprising at all: the most critical step in any mesh generation pipeline consists in placing additional points in the domain, because their coordinates must \emph{snap} to numbers that are representable in the floating point system, possibly introducing roundoff  errors. 
We resolved this issue using a modified version of earcut that prioritizes its ears according to their inner angle, thus producing better triangulations and ultimately permitting us to successfully process all the models in our dataset. We can conclude that with this tiny modification the algorithm is robust both in theory and in practice, without necessitating the additional overhead of exact point coordinates kernels~\cite{CLSA20}.


\section{Extension to Volumes}
\label{sec:3dmaps}
Besides its practical utility for 2D mappings, a major source of interest for this research deals with a volumetric extension of the algorithms proposed in Sections~\ref{sec:method} and~\ref{sec:extension}, in some way. In this section we analyze the major differences and challenges that we foresee.


\textbf{Convexity and Volumetric Earcut.}
In 2d the action of cutting an ear amounts to remove a convex vertex, substituting it with a straight segment that connects its two neighbors. Not only this operation is uniquely defined (there is one line passing through two given points), but it also decreases the inner angle of the neighbor vertices, preserving their convexity and, in turn, the convexity of the entire domain (if originally present). In 2d we deliberately exploit this property to restrict the quest for the next valid advancing move to one front only. For volumes, the situation is equivalent if and only if the convex vertex has valence 3. In such a case, cutting an ear amounts to delete the vertex, substituting it with a plane that interpolates its three neighbors, which is uniquely defined as well. This operation preserves the convexity of the domain, because dihedral angles can only decrease, and is the perfect dual of its 2d counterpart. Unfortunately, if the number of neighbors is higher than 3 (which is the most typical case for simplicial meshes) the operation of cutting an ear is not well defined, because the neighbor vertices may not be coplanar, and -- even if they were -- there would be multiple ways to triangulate them. One may still think to remove the vertex and triangulate the so generated pocket (some interesting ideas on how to tessellate these regions can be found in~\cite{chazelle1994bounds}), but if the neighbors are not coplanar some of the possible tessellations will not preserve convexity, which is a crucial property for the algorithm, and some other tessellations may not be applicable to the other domain, causing a deadlock. These observations suggest that domain convexity may be challenging to preserve throughout the mapping process, hence the ability to handle mappings to non convex domains must be devised.



\textbf{Deadlocks and Steiner Points.}
Another important difference between 2D and 3D mesh generation is the ability to design a tessellation without inserting additional points in the domain. This is always possible in 2D~\cite{meisters1975polygons}, but not in 3D. For the volumetric case vertex insertion is a classical problem, and indecomposable polyhedra that cannot be tetrahedralized without Steiner points were already known almost one century ago~\cite{schonhardt1928zerlegung}. 

Indeed, Steiner points could be used to tetrahedralized indecomposable input domains, and also to unlock deadlock configurations when both domains are non convex. However, 
this re-opens a number of classical computational geometry questions that were answered for the meshing of a single domain and -- to the best of our knowledge -- have no answer for the special case of compatible meshing of two domains at once. Specifically:

\begin{itemize}	
\item in case of a deadlock caused by two fronts that are indecomposable polyhedra, is it always possible to create a new move by adding one Steiner point in each domain? If not, what is the bound in the number of Steiner points that are necessary to grant the existence of a new valid move to advance both fronts simultaneously?\\

\item what if the deadlock is caused by the empty intersection between the valid moves in the two fronts, but considering each front alone a move exists? Is this case analogous to the point before?\\

\item what is the best positioning of a Steiner point in the context of simultaneous advancing front meshing? Can the coordinates of such point be expressed by rational numbers? This has huge practical importance, because it would guarantee that these points could be correctly positioned by a computer program (this is not the case for irrational numbers)\\

\item is there a bound on the global number of Steiner points necessary to simultaneously triangulate two polyhedra? For single volumes, we know from theory that this number is bounded by $\mathcal{O}(1)$ from below (see the Sch\"onhardt polyhedron~\cite{schonhardt1928zerlegung} and the subsequent generalization provided by Bagemihl~\cite{bagemihl1948indecomposable}), and by $\mathcal{O}(n^2)$ from above (by the Chazelle polyhedron~\cite{chazelle1984convex}). Should we expect similar bounds to exist also for pairs of shapes?\\
	
\item can we guarantee that compatible advancing front meshing always convergences?\\
\end{itemize}

Looking at previous literature for the meshing of a single domain we are tempted to be optimistic about the existence of reasonable (i.e. polynomial) bounds in the number of Steiner points, and the possibility to always unlock a deadlock configuration in $\mathcal{O}(1)$. Nevertheless, precise and theoretically sound answers to all these questions should be provided in order to grant robust tools for the generation of volumetric or non convex planar mappings.





\section{Conclusions}
\label{sec:conclusions}
We have proposed a novel algorithm to generate maps between simple polygons. At the core of this article is the proposal of a novel paradigm for the robust generation of simplicial maps, which is rooted in the principles of compatible mesh generation. We assume the input to be just a boundary representation of the domains to be connected, and we consider the mesh connectivity as an additional unknown. We then formulate the mapping as a mesh generation problem, where one wants to construct the same mesh in two different embeddings. This differentiates from classical approaches, which assume the topology of the mesh to be fixed, and solve the problem of positioning the interior vertices of a given mesh inside target domain. To grant maximal adoption we publicly released a reference implementation of our algorithms in Cinolib~\cite{cinolib}.

As witnessed by our results, the algorithms we propose are both theoretically sound and practically robust, and cover a wider range of mapping domains than prior methods, which are limited to convex~\cite{tutte1963draw,shen2019progressive}. On the other hand, our tools can only be used in a 2D-to-2D fashion, because a closed curve unambiguously identify a topological space only in 2d. Prior methods input a pre-existing surface mesh, which can be embedded in any higher dimensional space, hence~\cite{tutte1963draw,shen2019progressive} can also be used in a 3D-to-2D fashion.

We are mainly interested in this novel paradigm because it opens to a possible extension to volumetric meshes. This is a major difference w.r.t. prior art, because existing robust 2D approaches cannot extend to 3d (Section~\ref{sec:related}). To this end, in the final part of the paper we discussed potential issues that arise going one dimension up, and revived a number of classical computational geometry questions that are already answered for the meshing of a single domain, but have no answer for the novel case of compatible meshing. We believe that answering these questions will provide powerful tools to attack the volumetric mapping problem via mesh generation, and we will devote our future works to further develop this line of research. At the moment it is impossible to say whether a 3D extension would be computationally feasible, but the literature and wide availability of efficient tools for volume mesh generation leaves us reasonable hopes that in the near future a mapping method of this kind could be implemented.

\ifCLASSOPTIONcompsoc
  \section*{Acknowledgments}
\else
  \section*{Acknowledgment}
\fi
Thanks are due to Giorgio Luciano, for help with the procedural generation of 2D shapes in Blender.


\ifCLASSOPTIONcaptionsoff
  \newpage
\fi


\bibliographystyle{IEEEtran}
\bibliography{00_main}

\begin{IEEEbiography}[{\includegraphics[width=1in,height=1.25in,clip,keepaspectratio]{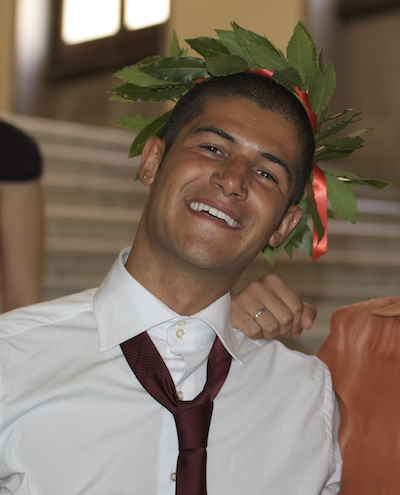}}]{Marco Livesu}
is a tenured researcher at the Institute for Applied Mathematics and Information Technologies of the National Research Council of Italy (CNR IMATI). He received his PhD at University of Cagliari in 2014, after which he was post doctoral researcher at the University of British Columbia, University of Cagliari and CNR IMATI. His main research interests are in computer graphics and geometry processing.
\end{IEEEbiography}



\end{document}